\documentclass[runningheads]{llncs}
\usepackage{graphicx}
\usepackage{amsmath,amssymb} 
\usepackage{color}
\usepackage{multirow}
\usepackage{booktabs}
\usepackage{xspace, xcolor}
\usepackage{subfigure}
\usepackage[export]{adjustbox}
\usepackage{url}

\def\eg{\textit{e.g.}}
\def\ie{\textit{i.e.}}
\def\etal{\textit{et al. }}

\definecolor{Gray}{gray}{0.9}
\begin{document}
\title{Frequency-Supervised\\MR-to-CT Image Synthesis}
%
%
\author{Zenglin Shi$^1$, Pascal Mettes$^1$, Guoyan Zheng$^2$, and Cees Snoek$^1$}
%
%
%
\institute{$^1$University of Amsterdam, $^2$Shanghai Jiao Tong University}
\maketitle              

\begin{abstract}
This paper strives to generate a synthetic computed tomography (CT) image from a magnetic resonance (MR) image. 
The synthetic CT image is valuable for radiotherapy planning when only an MR image is available.
Recent approaches have made large strides in solving this challenging synthesis problem with convolutional neural networks that learn a mapping from MR inputs to CT outputs. In this paper, we find that all existing approaches share a common limitation: reconstruction breaks down in and around the high-frequency parts of CT images. To address this common limitation, we introduce frequency-supervised deep networks to explicitly enhance high-frequency MR-to-CT image reconstruction. We propose a frequency decomposition layer that learns to decompose predicted CT outputs into low- and high-frequency components, and we introduce a refinement module to improve high-frequency reconstruction through high-frequency adversarial learning.
Experimental results on a new dataset with 45 pairs of 3D MR-CT brain images show the effectiveness and potential of the proposed approach. Code is available at \url{https://github.com/shizenglin/Frequency-Supervised-MR-to-CT-Image-Synthesis}.

\end{abstract}

\section{Introduction}
Magnetic resonance (MR) image is widely used in clinical diagnosis and cancer monitoring, as it is obtained through a non-invasive imaging protocol, and it delivers excellent soft-tissue contrast. However, MR image does not provide electron density information that computed tomography (CT) image can provide, which is essential for applications like dose calculation in radiotherapy treatment planning \cite{Burgos_2017,guerreiro2017evaluation,burgos2017iterative,largent2019pseudo} and attenuation correction in positron emission tomography reconstruction \cite{Navalpakkam_IR_2013,Liu_Radiology_2017,klaser2019improved}. 
To overcome this limitation, a variety of approaches have been proposed to recreate a CT image from the available MR images \cite{Nie2016Estimating,Han_MP_2017,Nie2017Medical,wolterink2017mr,ge2019unpaired}. Recently, deep learning-based synthesis methods \cite{Nie2016Estimating,Han_MP_2017,Nie2017Medical,wolterink2017mr,ge2019unpaired,largent2019comparison} have shown superior performance over alternatives such as segmentation-based \cite{Korhonen_MP_2014,berker2012mri,burgos2017iterative} and atlas-based methods \cite{sjolund2015generating,burgos2015robust,degen2016multi,burgos2016joint}. 

A typical approach for deep learning-based synthesis is through 2D convolutional networks on 2D MR images~\cite{ge2019unpaired,Han_MP_2017,ronneberger2015u,wolterink2017mr,largent2019pseudo,emami2018generating,jin2019deep}. A downside of this setup is that 2D approaches are applied to 3D MR images slice-by-slice, which can cause discontinuous prediction results across slices \cite{Nie2017Medical}. To take full use of 3D spatial information of volumetric data, 3D-based synthesis models have been explored using 3D convolutional networks \cite{Nie2016Estimating,roy2017synthesizing} and 3D GANs \cite{Nie2017Medical}. In this work we adopt a similar setup, which uses paired MR and CT images during training, but we tackle a common limitation amongst existing 3D-based synthesis approaches: imperfect CT image synthesis in high-frequency parts of the volume.

\begin{figure}[t!]
 \centering
		\includegraphics[width=\textwidth]{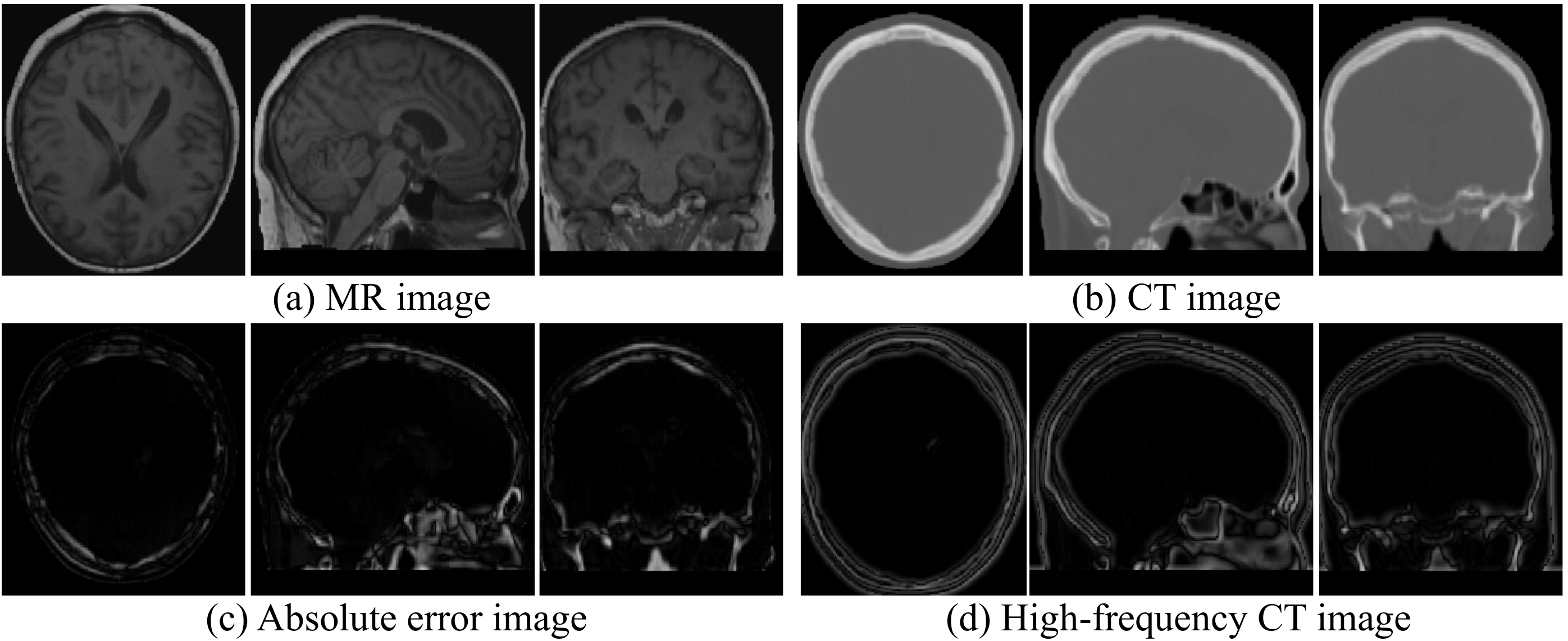}
		\caption{
		\textbf{High-frequency supervision motivation} for MR-to-CT image synthesis. For MR (a) to CT (b) image synthesis using 3D networks \cite{3DUNET_2016}, the reconstruction error (c) is most dominant in regions that directly overlap with the high-frequency parts of the CT image (d). For computing the high-frequency CT image, we first obtain a low-frequency CT image through a Gaussian low-pass filter. Then, we subtract the low-frequency CT image from the raw CT image to generate a high-frequency CT image. In the error image (c), the brighter the voxel, the bigger the error. 
		} 
\label{fig_fig1}
\end{figure}

The main motivation behind our work is visualized in Fig.~\ref{fig_fig1}. For MR (a) to CT (b) image synthesis using 3D networks \cite{3DUNET_2016}, the reconstruction error (c) is most dominant in regions that directly overlap with the high-frequency parts of the CT image (d). This is a direct result of the used loss function, \eg, an $\ell_1$ or $\ell_2$ loss, which results in blurring since they are minimized by averaging all possible outputs~\cite{isola2017image,mathieu2015deep}. As a result, the low-frequency parts are reconstructed well, at the cost of the high-frequency parts. Interestingly, Lin \etal \cite{lin2020frequency} also found CNN-based synthesis models tend to lose high-frequency image details for CT-to-MR image synthesis. To address this limitation, they propose the frequency-selective learning, where multiheads are used in the deep network for learning the reconstruction of different frequency components. Differently, in this work, we propose frequency-supervised networks that explicitly aim to enhance high-frequency reconstruction in MR-to-CT image synthesis.

We make three contributions in this work: \textit{i)} we propose a frequency decomposition layer to decompose the predicted CT image into high-frequency and low-frequency parts. This decomposition is supervised by decomposing ground truth CT images using low-pass filters. In this way, we can focus on improving the quality of the high-frequency part, assisted by \textit{ii)} a high-frequency refinement module. This module is implemented as a 3D symmetric factorization convolutional block to maximize reconstruction performance with minimal parameters; and \textit{iii)} we outline a high-frequency adversarial learning to further improve the quality of the high-frequency CT image. Experimental results on a dataset with 45 pairs of 3D MR-CT brain images shows the effectiveness and potential of the proposed approach.

\section{Method}
We formulate the MR-to-CT image synthesis task as a 3D image-to-image translation problem. Let $ \mathcal{X} = {\{x_i}\}_{i=1}^{H \times W \times L}$ be an input MR image of size ${H \times W \times L}$, and $\mathcal{Y} = {\{y_i}\}_{i=1}^{H \times W \times L}$ be the target CT image for this MR image, where $y_i$ is the target voxel for the voxel of $x_i$. The transformation from input images to target images can be achieved by learning a mapping function, \ie, $f: \mathcal{X} \mapsto \mathcal{Y}$. In this paper, we learn the mapping function by way of voxel-wise nonlinear regression, implemented through a 3D convolutional neural network. Let $\Psi(\mathcal{X}): \mathbb{R}^{H \times W \times L} \mapsto \mathbb{R}^{H \times W \times L}$ denote such a mapping given an arbitrary 3D convolutional neural network $\Psi$ for input MR image $\mathcal{X}$.
\begin{figure}[t]
\centering 
\includegraphics[width=\linewidth,left]{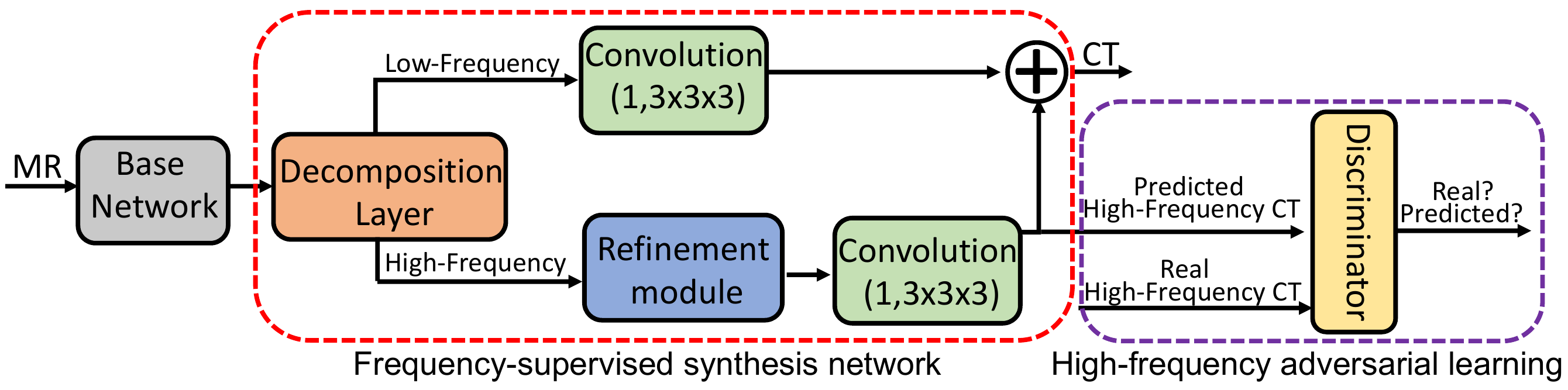}
\caption{\textbf{Frequency-supervised network architecture}. Our approach is agnostic to the base 3D MR-to-CT image synthesis network. The decomposition layer splits the output features of the 3D base network into two parts that generate the low-frequency and high-frequency components of the CT image. Then a refinement module improves synthesis in the high-frequency parts and is explicitly learned with a specific high-frequency supervision. Finally, the predicted high-frequency CT image is further enhanced by means of adversarial learning. $\oplus$ denotes the element-wise sum.}
\label{fig:fig_network}  
\end{figure}
\\\\
\subsection{Frequency-supervised synthesis network}
We propose a 3D network that specifically emphasizes the high-frequency parts of CT images. Standard losses for 3D networks, such as the $\ell_1$ loss, perform well in the low-frequency parts of CT images, at the cost of loss in precision for the high-frequency parts. Our approach is agnostic to the base 3D network and introduces two additional components to address our desire for improved synthesis in the high-frequency parts: a decomposition layer and a refinement module, which is explicitly learned with a specific high-frequency supervision. The overall network is visualized in Figure~\ref{fig:fig_network}.
\\\\
\textbf{Decomposition layer.} We account for a specific focus on high-frequency CT image parts through a decomposition layer. This layer learns to split the output features of a 3D base network into low-frequency and high-frequency components in a differentiable manner. Let $V=\Psi(\mathcal{X}) \in \mathbb{R}^{C\times H \times W \times L}$ denote the output of the penultimate layer of the base network for input $\mathcal{X}$. We add a $3 \times 3 \times 3$ convolution layer with parameters $\theta_d\in \mathbb{R}^{C\times 2 \times 3 \times 3 \times 3}$, followed by a softmax function to generate probability maps $P=[p_l,p_h]=\text{softmax}(\theta_dV) \in \mathbb{R}^{2\times H \times W \times L}$. The probability scores for each voxel denote the likelihood of the voxel belonging to the low- or high-frequency parts of the CT images. Using the probabilities, we obtain low-frequency features $V_l=p_l*V$ and high-frequency features $V_h=p_h*V$, where $p_l$ and $p_h$ are first tiled to be the same size as $V$. For the low-frequency part, we use a $3 \times 3 \times 3$ convolution layer with the parameters $\theta_l\in \mathbb{R}^{C\times 1 \times 3 \times  \times 3}$ to generate the low-frequency CT image $\hat{\mathcal{Y}}_l=\theta_lV_l$.
\\\\
\textbf{Refinement module.}
To generate the high-frequency CT image $\hat{\mathcal{Y}}_h$, we introduce a refinement module to improve the quality of the high-frequency features $V_h$. Since the high-frequency features are close to zero in most regions, its learning usually requires a large receptive field for capturing enough context information. The enhancement is performed on top of a base network, \eg, a 3D U-Net \cite{3DUNET_2016}, thus we should limit the amount of extra parameters and layers to avoid making the network hard to optimize. To this end, we introduce a 3D symmetric factorization module. The module explicitly factorizes a 3D convolution into three 1D convolutions along three dimensions. To process each dimension equally, the module employs a symmetric structure with the combination of ($k \times 1 \times 1$)+($1 \times k \times 1$)+($1 \times 1 \times k$), ($1 \times k \times 1$)+($1 \times 1 \times k$)+($k \times 1 \times 1$), and ($1 \times 1 \times k$)+($k \times 1 \times 1$)+($1 \times k \times 1$) convolutions. Specifically, the input of this module is convolved with three 1D convolutions for each dimension, the output of each 1D convolution is convolved two more times over the remaining dimensions. Then the outputs of last three 1D convolutions is summed as the output of this module. Compared to a standard 3D $k \times k \times k$ convolution layer, parameters is reduced from $k^3$ to $9k$. In this paper, we use $k=13$ for relatively large receptive field. The module is denoted as $\phi$ with the parameters $\theta_e$. Then we use a $3 \times 3 \times 3$ convolution layer with the parameters $\theta_h\in \mathbb{R}^{C\times 1 \times 3 \times 3 \times 3}$ to generate the high-frequency CT image $\hat{\mathcal{Y}}_h=\theta_h\phi(V_h)$. 
\\\\
\textbf{Optimization.}
We use a specific loss for high-frequency CT image synthesis to explicitly learn the high-frequency refinement module. Another loss is used for overall CT image synthesis. Empirically, low-frequency CT image can be synthesised correctly with only an overall loss. Thus, we minimize the difference between predicted high-frequency CT image $\hat{\mathcal{Y}_h}$ and its ground-truth $\mathcal{Y}_h$, and the difference between predicted CT image $(\hat{\mathcal{Y}_h}+\hat{\mathcal{Y}_l})$ and its ground-truth $\mathcal{Y}$ using the $L_1$-norm, which is defined as:
\begin{equation}
\mathcal{L} = \parallel\hat{\mathcal{Y}}_h-\mathcal{Y}_h \parallel_{1}+\parallel (\hat{\mathcal{Y}}_l+\hat{\mathcal{Y}}_h) - \mathcal{Y}\parallel_{1}.
\label{eq_genc}
\end{equation}

During training, the ground truth high-frequency CT image $\mathcal{Y}_h$ is obtained fully automatically without \textit{any} manual labeling. Specifically, we first obtain a low-frequency CT image through a Gaussian low-pass filter with filtering size $\sigma=15$. Then we subtract the low-frequency CT image from the CT image to obtain the high-frequency CT image. During inference, we input a MR $\mathcal{X}$, and output $(\hat{\mathcal{Y}}_l+\hat{\mathcal{Y}}_h)$ as the synthesised CT image.

\subsection{High-frequency adversarial learning}
\label{sec_gan}
Lastly, we enhance the predicted high-frequency CT image $\hat{\mathcal{Y}}_h$ by means of adversarial learning. Adversarial learning has shown its benefits in MR-to-CT image synthesis by Nie \etal \cite{Nie2017Medical}. We have observed that the low-frequency CT image can be reconstructed well. Thus, we propose to apply the discriminator only on the high frequencies. This reduces the complexity of the problem, making it easier for the discriminator to focus on the relevant image features. We use the relativistic discriminator introduced by \cite{jolicoeur2018relativistic}. The discriminator makes adversarial learning considerably more stable and generates higher quality images.

\section{Experiments and results}
\subsection{Experimental setup} 
\textbf{Dataset and pre-processing.} We evaluate our approach on a dataset with 45 pairs of 3D MR-CT brain images.  When comparing the size of our data to previous supervised works, our data set size is reasonable. Such as, Nie \etal \cite{Nie2017Medical} report on 38 data pairs and Han \etal \cite{Han_MP_2017} use 33 data pairs. Our images are acquired in the head region for the clinical indications of dementia, epilepsy and grading of brain tumours. The MR images have a spacing of $0.8 \times 0.8 \times 0.8$ $mm^3$ while the CT images have a spacing of $0.9 \times 0.9 \times 2.5$ $mm^3$. Registration is performed to align the two modalities and to sample the aligned images with a spacing of $1.0\times1.0\times1.5 \ mm^3 $. The gray values of the CT were uniformly distributed in the range of $[-1024$, $2252.7]$ Hounsfield unit. We resample all the training data to isotropic resolution and normalized each MR image as zero mean and unit variance. We also normalize each CT image into the range of $[0,1]$ and we expect that the output synthetic values are also in the same range. The final synthetic CT will be obtained by multiplying the normalized output with the range of Hounsfield unit in our training data.
\\\\
\textbf{Implementation details.}
Implementation is done with Python using TensorFlow. 
Network parameters are initialized with He initialization and trained using Adam with a mini-batch of 1 for 5,000 epochs. We set $\beta_1$ to 0.9, $\beta_2$ to 0.999 and the initial learning rate to 0.001. Data augmentation is used to enlarge the training samples by rotating each image with a random angle in the range of $[-10^\circ, 10^\circ]$ around the $z$ axis. Randomly cropped $64 \times 64 \times 64$ sub-volumes serve as input to train our network.  During testing, we adopt sliding window and overlap-tiling stitching strategies to generate predictions for the whole volume. We use MAE (Mean Absolute Error), PSNR (Peak Signal-to-Noise Ratio) and SSIM (Structural Similarity Index Measure) as evaluation measurements. These measurements are reported by 5-fold cross-validation. 

\subsection{Results}
\begin{figure}[t!]
 \centering
		\includegraphics[width=\textwidth]{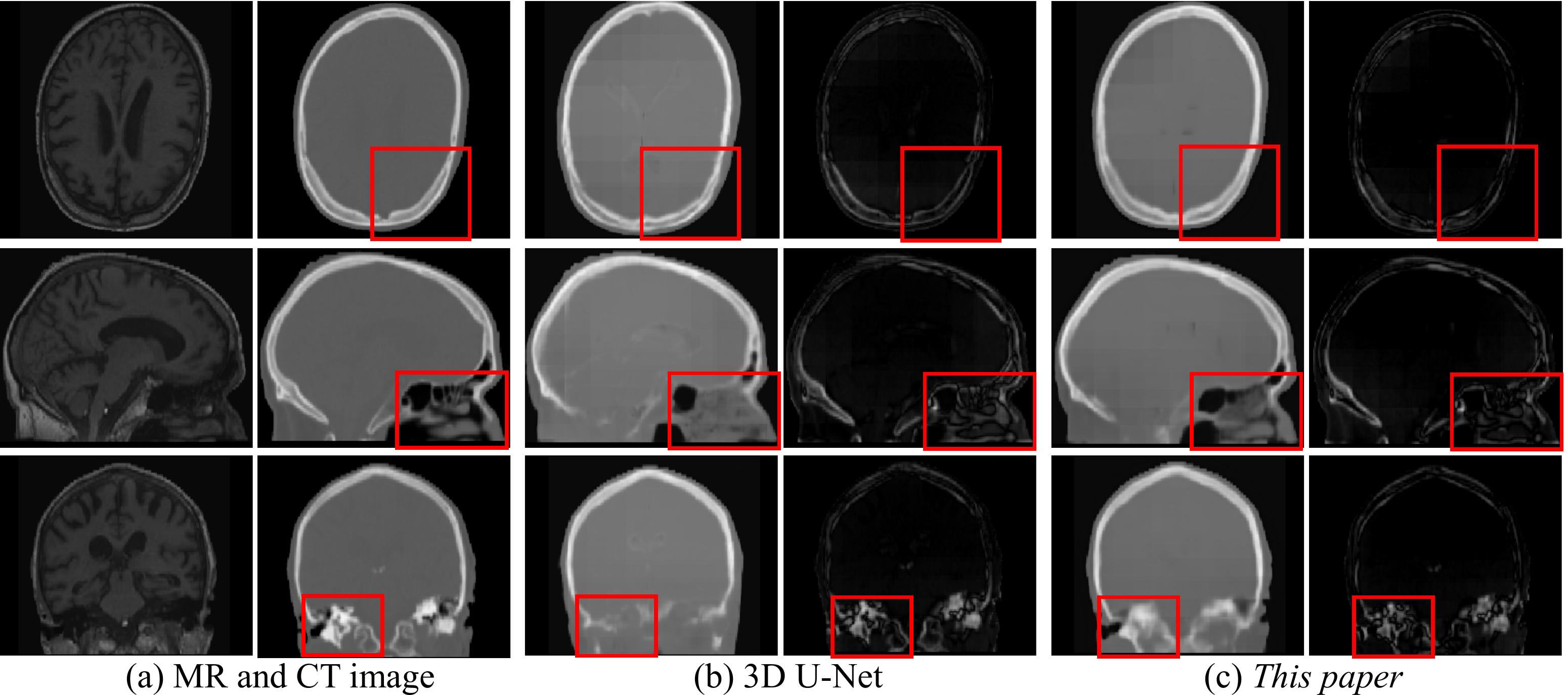}
		\caption{\textbf{Effect of frequency-supervised learning}. The right images of (b) and (c) are the error images, where the brighter the voxel, the bigger the error. Our method better reduces the synthesised errors than 3D U-Net, as highlighted by the red squares.}
\label{fig_decom}
\end{figure}

\begin{figure}[tbp]
 \centering
		\includegraphics[width=\textwidth]{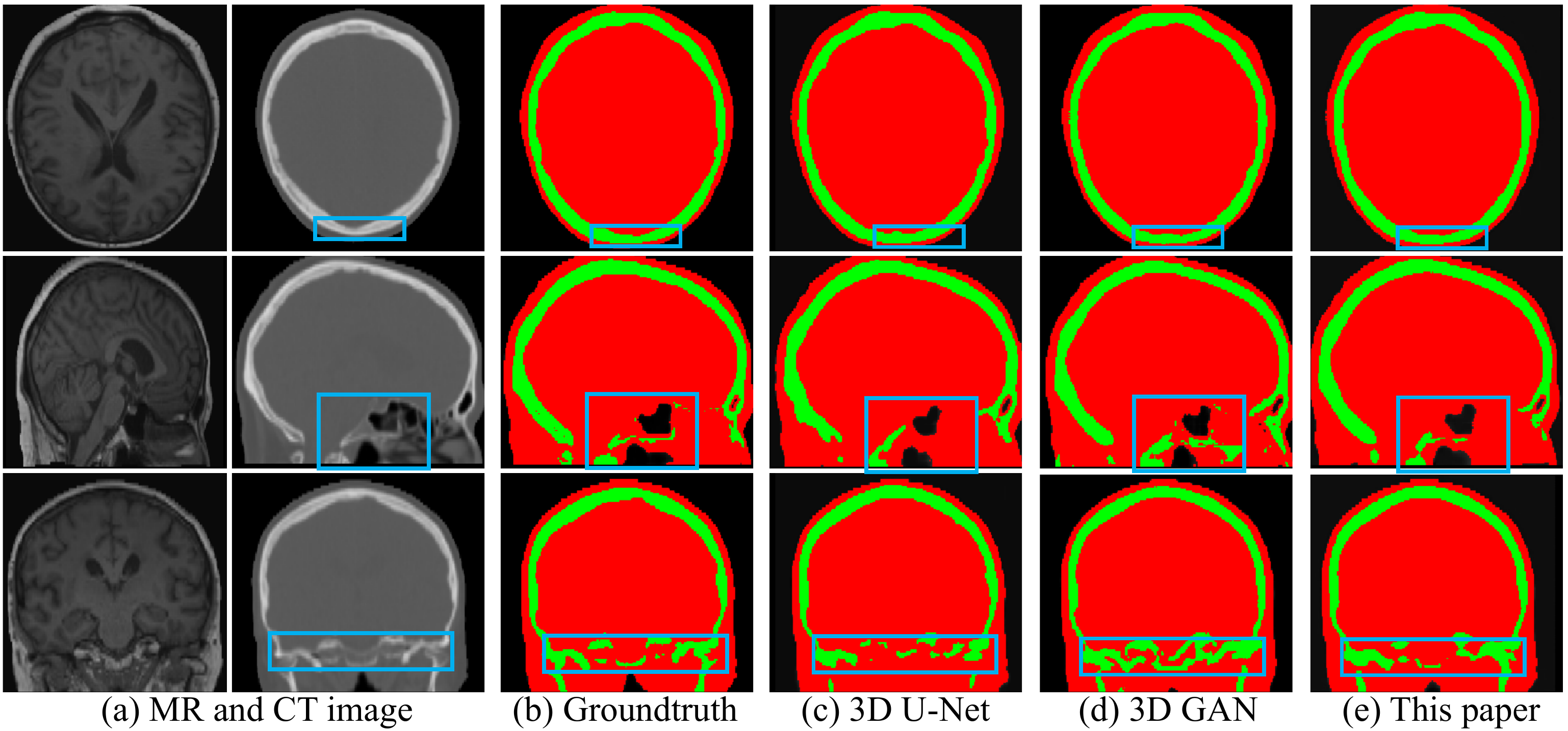}
		\caption{\textbf{Evaluations by segmentation}. The regions with black color represent background and air, the regions with red color represent soft tissue, the regions with green color represent bone tissue. Our method achieves better synthetic bone tissues, as highlighted by the blue squares.}
\label{fig_seg}
\end{figure}

\textbf{Effect of frequency-supervised learning.}
We first analyze the effect of the proposed frequency-supervised learning. We compare it to two baselines. The first performs synthesis using the base network only. Here, we compare three widely used network architectures, \ie, 3D fully convolutional network (3D FC-Net) \cite{Nie2016Estimating}, 3D residual network (3D Res-Net) \cite{li2017compactness} and 3D U-Net \cite{3DUNET_2016}. 
The second baseline adds a boundary refinement module introduced by~\cite{peng2017large} on top of the base network, which improves the structures of the predicted image by means of residual learning. The losses of these two baseline models are optimized with respect to the overall CT image estimation. As a result, low-frequency parts of predicted CT image are reconstructed well, at the cost of high-frequency parts. By contrast, we first decompose the predicted CT image into low-frequency and high-frequency parts, and then improve the quality of high-frequency parts with a high-frequency refinement module, which is learned by a specific high-frequency loss. The results are shown in Table~\ref{tab_fsup}. 
With the base network only, 3D U-Net obtains the best performance. With the addition of the boundary refinement module, only the performance of the 3D FC-Net is improved. By contrast, our method improves over all three base networks. Fig. \ref{fig_decom} highlights our ability to better reduce synthesised errors, especially in high-frequency parts. We conclude our frequency-supervised learning helps MR-to-CT image synthesis, regardless of the 3D base network, and we will use 3D U-Net as basis for further experiments.
\begin{table}[t]
\small
\caption{\textbf{Effect of frequency-supervised learning.} For all base networks, our high-frequency supervised learning results in improved synthesis across all metrics.}
\centering
\resizebox{\linewidth}{!}{
\begin{tabular}{@{}lcccccccccc@{}}
\toprule
 & \multicolumn{3}{c}{\textbf{3D FC-Net} \cite{Nie2016Estimating}} & \multicolumn{3}{c}{\textbf{3D Res-Net} \cite{li2017compactness}} & \multicolumn{3}{c}{\textbf{3D U-Net} \cite{3DUNET_2016}}\\
\cmidrule(lr){2-4} \cmidrule(lr){5-7} \cmidrule(lr){8-10}
& MAE$\downarrow$ & PSNR$\uparrow$ & SSIM$\uparrow$ & MAE$\downarrow$ & PSNR$\uparrow$ & SSIM$\uparrow$ & MAE$\downarrow$ & PSNR$\uparrow$ & SSIM$\uparrow$\\
\hline
Base network & 94.55  &  24.43 & 0.681 & 81.26  &  25.89 & 0.724 & 79.09  &  26.10 & 0.726  \\
\textbf{w/} Boundary refinement \cite{peng2017large} & 90.15  &  25.03 & 0.697 & 82.54  &  25.76 & 0.723& 82.83  &  25.63 & 0.723  \\
\textbf{w/} \textbf{\textit{This paper}} & \textbf{84.31}  &  \textbf{25.72} & \textbf{0.736} & \textbf{73.61}  &  \textbf{26.72} & \textbf{0.741} & \textbf{72.71} & \textbf{26.86} & \textbf{0.747}  \\
\bottomrule
\end{tabular}}
\label{tab_fsup}
\end{table}
\begin{table}[t]
\small
\caption{\textbf{Effect of refinement module.} $L$ denotes layer number, $C$ denotes channel number and $K$ denotes kernel size.  Our proposed refinement module introduces relatively few parameters with large receptive fields, leading to improved performance. }
\centering
\resizebox{0.98\linewidth}{!}{
\begin{tabular}{@{}lcclcclcc@{}}
\toprule
 \multicolumn{3}{c}{\textbf{Stacking}} & \multicolumn{3}{c}{\textbf{Large kernel size}} & \multicolumn{3}{c}{\textbf{This paper}}\\
\cmidrule(lr){1-3} \cmidrule(lr){4-6} \cmidrule(lr){7-9}
3D Convolutions & MAE$\downarrow$ & \#params &
3D Convolutions & MAE$\downarrow$ & \#params & 
1D Convolutions & MAE$\downarrow$ & \#params\\
\hline
($L=3$, $C=32$, $K=3$)  & 76.38  &  82.9k &
($L=3$, $C=32$, $K=3$)  & 76.38  &  82.9k &
($L=9$, $C=32$, $K=3$)  & 74.08  &  64.5k  \\
($L=6$, $C=32$, $K=3$)  & 79.60  &  165.9k &
($L=3$, $C=32$, $K=5$)  & 75.78  &  384.0k &
($L=9$, $C=32$, $K=13$)  & 72.71  &  119.8k  \\
($L=9$, $C=32$, $K=3$)  & 81.89  &  248.8k & 
($L=3$, $C=32$, $K=7$)& 79.09  &  1053.7k &
($L=9$, $C=32$, $K=19$)  & 72.69  &  175.1k \\
\bottomrule
\end{tabular}
}
\label{tab_enha}
\end{table}
\\\\
\textbf{Effect of high-frequency refinement module.} 
In the second experiment, we demonstrate the effect of the proposed refinement module. To capture more context information by enlarging the receptive field, standard approaches include stacking multiple $3 \times 3 \times 3$ convolutional layers or use convolutions with larger kernel, \eg, $7 \times 7 \times 7$. The experimental results in Table \ref{tab_enha} show that neither approach works well. Such as, the MAE of stacking three $3 \times 3 \times 3$ convolutional layers is 76.38, while the MAE of stacking six is 79.60. The MAE of stacking three $3 \times 3 \times 3$ convolutional layers is 76.38, while the MAE of stacking three $7 \times 7 \times 7$ is 79.09. By contrast, our proposed module with $k=13$ introduces relatively few parameters with large convolution kernels, leading to a better MAE of 72.71.
\begin{table}[t!]
\small
\caption{\textbf{Effect of high-frequency adversarial learning.} Our approach, with both frequency-supervision and high-frequency adversarial learning, outperforms 3D U-Net with standard adversarial learning.}
\centering
\resizebox{0.99\columnwidth}{!}
{\begin{tabular}{@{}lccccc@{}}
\toprule
& MAE$\downarrow$ & PSNR$\uparrow$ & SSIM$\uparrow$ \\
\hline
3D U-Net  & 79.09  &  26.10 & 0.726 \\
3D GAN & 76.83  &  26.55 & 0.742 \\
\textit{This paper:} Frequency-supervised synthesis  & 72.71  &  26.86 & 0.747\\
\textit{\textbf{This paper: Frequency-supervised synthesis and adversarial learning}} & \textbf{69.57}  &  \textbf{27.39}&  \textbf{0.758} \\
\bottomrule
\end{tabular}
}
\label{tab_gan}
\end{table}
\begin{figure}[t!]
 \centering
		\includegraphics[width=\textwidth]{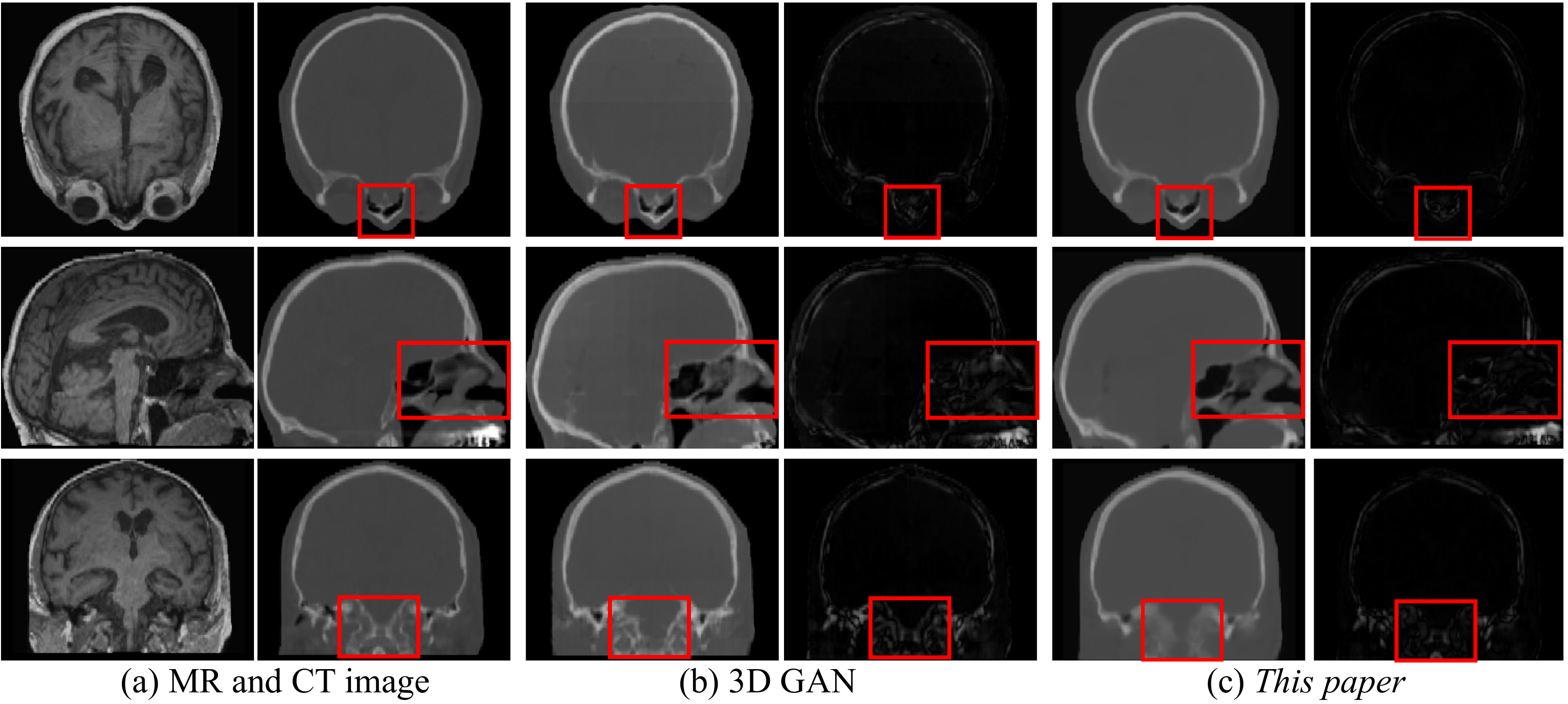}
		\caption{\textbf{Effect of high-frequency adversarial learning}. From the figure, one can observe that the proposed method reduces the synthesized error and yields synthesized CT images with better perceptive quality, as highlighted by the red squares. In the error images, the brighter the voxel, the bigger the error. }
\label{fig_gan}
\end{figure}
\\\\
\textbf{Effect of high-frequency adversarial learning.}
In this experiment, we show the effectiveness of the proposed high-frequency adversarial learning. For the generator network, we use the 3D U-Net for standard adversarial learning, and frequency-supervised synthesis network for our high-frequency adversarial learning, as shown in Fig. \ref{fig:fig_network}, where 3D U-Net is the base network. For the discriminator, we both use the relativistic average discriminator introduced in \cite{jolicoeur2018relativistic} (See section \ref{sec_gan}). The network architecture of the discriminator is the same as the encoder of 3D U-Net. 3D U-Net combined with standard adversarial learning leads to a 3D GAN based synthesis model, as the work of Nie \etal \cite{Nie2017Medical}. As shown in Table \ref{tab_gan}, the 3D GAN achieves better synthesis performance than 3D U-Net only. Our high-frequency adversarial learning further improves the performance of our frequency-supervised synthesis network. From Fig.~\ref{fig_gan}, we observe the proposed method yields synthesized CT images with better perceptive quality, in particular higher structural similarity and more anatomical details.
\\\\
\textbf{Evaluation by segmentation}
To further evaluate the quality of synthesised CT images generated by various methods. Following  Hangartner \cite{hangartner2007thresholding}, we use a simple thresholding to segment ground-truth and synthesised CT images into three classes: 1) background and air; 2) soft tissue; and 3) bone tissue. As shown in Fig. \ref{fig_seg}, our proposed methods mainly improve the model's capability on bone tissue synthesis compared to other methods.

\section{Conclusion}
In this paper, we have shown that existing deep learning based MR-to-CT image synthesis methods suffer from high-frequency information loss in the synthesized CT image. To enhance the reconstruction of high-frequency CT images, we present a frequency-supervised MR-to-CT image synthesis method. Our method contributes a frequency decomposition layer, a high-frequency refinement module and a high-frequency adversarial learning, which are combined to explicitly improve the quality of the high-frequency CT image. Our experimental results demonstrate the effectiveness of the proposed methods.

%
%
\label{sect:bib}
\bibliographystyle{splncs04}
\bibliography{ref}

\begin{thebibliography}{10}
\providecommand{\url}[1]{\texttt{#1}}
\providecommand{\urlprefix}{URL }
\providecommand{\doi}[1]{https://doi.org/#1}

\bibitem{berker2012mri}
Berker, Y., Franke, J., Salomon, A., Palmowski, M., Donker, H.C., Temur, Y.,
  Mottaghy, F.M., Kuhl, C., Izquierdo-Garcia, D., Fayad, Z.A., et~al.:
  Mri-based attenuation correction for hybrid pet/mri systems: a 4-class tissue
  segmentation technique using a combined ultrashort-echo-time/dixon mri
  sequence. Journal of nuclear medicine  \textbf{53}(5),  796--804 (2012)

\bibitem{Burgos_2017}
Burgos, N., Guerreiro, F., et~al.: Iterative framework for the joint
  segmentation and ct synthesis of mr images: application to mri-only
  radiotherapy treatment planning. Phys Med Biol  \textbf{62},  4237--4253
  (2017)

\bibitem{burgos2015robust}
Burgos, N., Cardoso, M.J., Guerreiro, F., Veiga, C., Modat, M., McClelland, J.,
  Knopf, A.C., Punwani, S., Atkinson, D., Arridge, S.R., et~al.: Robust ct
  synthesis for radiotherapy planning: application to the head and neck region.
  In: MICCAI. pp. 476--484 (2015)

\bibitem{burgos2016joint}
Burgos, N., Guerreiro, F., McClelland, J., Nill, S., Dearnaley, D., Desouza,
  N., Oelfke, U., Knopf, A.C., Ourselin, S., Cardoso, M.J.: Joint segmentation
  and ct synthesis for mri-only radiotherapy treatment planning. In: MICCAI.
  pp. 547--555 (2016)

\bibitem{burgos2017iterative}
Burgos, N., Guerreiro, F., McClelland, J., Presles, B., Modat, M., Nill, S.,
  Dearnaley, D., Desouza, N., Oelfke, U., Knopf, A.C., et~al.: Iterative
  framework for the joint segmentation and ct synthesis of mr images:
  application to mri-only radiotherapy treatment planning. Physics in Medicine
  \& Biology  \textbf{62}(11), ~4237 (2017)

\bibitem{3DUNET_2016}
{\c{C}}i{\c{c}}ek, {\"O}., Abdulkadir, A., et~al.: 3d u-net: learning dense
  volumetric segmentation from sparse annotation. In: MICCAI. vol.~9901, pp.
  424--432 (2016)

\bibitem{degen2016multi}
Degen, J., Heinrich, M.P.: Multi-atlas based pseudo-ct synthesis using
  multimodal image registration and local atlas fusion strategies. In: CVPR
  Workshops. pp. 160--168 (2016)

\bibitem{emami2018generating}
Emami, H., Dong, M., Nejad-Davarani, S.P., Glide-Hurst, C.K.: Generating
  synthetic cts from magnetic resonance images using generative adversarial
  networks. Medical physics  \textbf{45}(8),  3627--3636 (2018)

\bibitem{ge2019unpaired}
Ge, Y., Wei, D., Xue, Z., Wang, Q., Zhou, X., Zhan, Y., Liao, S.: Unpaired mr
  to ct synthesis with explicit structural constrained adversarial learning.
  In: ISBI (2019)

\bibitem{guerreiro2017evaluation}
Guerreiro, F., Burgos, N., Dunlop, A., Wong, K., Petkar, I., Nutting, C.,
  Harrington, K., Bhide, S., Newbold, K., Dearnaley, D., et~al.: Evaluation of
  a multi-atlas ct synthesis approach for mri-only radiotherapy treatment
  planning. Physica Medica  \textbf{35},  7--17 (2017)

\bibitem{Han_MP_2017}
Han, X.: Mr-based synthetic ct generation using a deep convolutional neural
  network. Med. Phys.  \textbf{44}(4),  1408--1419 (2017)

\bibitem{hangartner2007thresholding}
Hangartner, T.N.: Thresholding technique for accurate analysis of density and
  geometry in qct, pqct and muct images. Journal of Musculoskeletal and
  Neuronal Interactions  \textbf{7}(1), ~9 (2007)

\bibitem{isola2017image}
Isola, P., Zhu, J.Y., Zhou, T., Efros, A.A.: Image-to-image translation with
  conditional adversarial networks. In: CVPR (2017)

\bibitem{jin2019deep}
Jin, C.B., Kim, H., Liu, M., Jung, W., Joo, S., Park, E., Ahn, Y.S., Han, I.H.,
  Lee, J.I., Cui, X.: Deep ct to mr synthesis using paired and unpaired data.
  Sensors  \textbf{19}(10), ~2361 (2019)

\bibitem{jolicoeur2018relativistic}
Jolicoeur-Martineau, A.: The relativistic discriminator: a key element missing
  from standard gan. In: ICLR (2019)

\bibitem{klaser2019improved}
Kl{\"a}ser, K., Varsavsky, T., Markiewicz, P., Vercauteren, T., Atkinson, D.,
  Thielemans, K., Hutton, B., Cardoso, M.J., Ourselin, S.: Improved mr to ct
  synthesis for pet/mr attenuation correction using imitation learning. In:
  SSMI Workshop. pp. 13--21 (2019)

\bibitem{Korhonen_MP_2014}
Korhonen, J., Kapanen, M., et~al.: A dual model hu conversion from mri
  intensity values within and outside of bone segment for mri-based
  radiotherapy treatment planning of prostate cancer. Med Phys  \textbf{41},
  011704 (2014)

\bibitem{largent2019pseudo}
Largent, A., Barateau, A., Nunes, J.C., Lafond, C., Greer, P.B., Dowling, J.A.,
  Saint-Jalmes, H., Acosta, O., de~Crevoisier, R.: Pseudo-ct generation for
  mri-only radiation therapy treatment planning: comparison among patch-based,
  atlas-based, and bulk density methods. International Journal of Radiation
  Oncology* Biology* Physics  \textbf{103}(2),  479--490 (2019)

\bibitem{largent2019comparison}
Largent, A., Barateau, A., Nunes, J.C., Mylona, E., Castelli, J., Lafond, C.,
  Greer, P.B., Dowling, J.A., Baxter, J., Saint-Jalmes, H., et~al.: Comparison
  of deep learning-based and patch-based methods for pseudo-ct generation in
  mri-based prostate dose planning. International Journal of Radiation
  Oncology* Biology* Physics  \textbf{105}(5),  1137--1150 (2019)

\bibitem{li2017compactness}
Li, W., Wang, G., Fidon, L., Ourselin, S., Cardoso, M.J., Vercauteren, T.: On
  the compactness, efficiency, and representation of 3d convolutional networks:
  brain parcellation as a pretext task. In: IPMI. pp. 348--360 (2017)

\bibitem{lin2020frequency}
Lin, Z., Zhong, M., Zeng, X., Ye, C.: Frequency-selective learning for ct to mr
  synthesis. In: International Workshop on Simulation and Synthesis in Medical
  Imaging. pp. 101--109. Springer (2020)

\bibitem{Liu_Radiology_2017}
Liu, F., Jang, H., et~al.: Deep learning mr imaging-based attenuation
  correction for pet/mr imaging. Radiology  \textbf{286}(2),  676--684 (2017)

\bibitem{mathieu2015deep}
Mathieu, M., Couprie, C., LeCun, Y.: Deep multi-scale video prediction beyond
  mean square error. ICLR  (2016)

\bibitem{Navalpakkam_IR_2013}
Navalpakkam, B., Braun, H., et~al.: Magnetic resonance-based attenuation
  correction for pet/mr hybrid imaging using continuous valued attenuation
  maps. Invest Radiol  \textbf{48},  323--332 (2013)

\bibitem{Nie2016Estimating}
Nie, D., Cao, X., et~al.: Estimating ct image from mri data using 3d fully
  convolutional networks. In: DLMIA. pp. 170--178 (2016)

\bibitem{Nie2017Medical}
Nie, D., Trullo, R., et~al.: Medical image synthesis with context-aware
  generative adversarial networks. In: MICCAI. pp. 417--425 (2017)

\bibitem{peng2017large}
Peng, C., Zhang, X., Yu, G., Luo, G., Sun, J.: Large kernel matters--improve
  semantic segmentation by global convolutional network. In: CVPR (2017)

\bibitem{ronneberger2015u}
Ronneberger, O., Fischer, P., Brox, T.: U-net: Convolutional networks for
  biomedical image segmentation. In: MICCAI (2015)

\bibitem{roy2017synthesizing}
Roy, S., Butman, J.A., Pham, D.L.: Synthesizing ct from ultrashort echo-time mr
  images via convolutional neural networks. In: SSMI Workshop. pp. 24--32
  (2017)

\bibitem{sjolund2015generating}
Sj{\"o}lund, J., Forsberg, D., Andersson, M., Knutsson, H.: Generating patient
  specific pseudo-ct of the head from mr using atlas-based regression. Physics
  in Medicine \& Biology  \textbf{60}(2), ~825 (2015)

\bibitem{wolterink2017mr}
Wolterink, J.M., Dinkla, A.M., Savenije, M.H., Seevinck, P.R., van~den Berg,
  C.A., I{\v{s}}gum, I.: Mr-to-ct synthesis using cycle-consistent generative
  adversarial networks. In: NIPS (2017)

\end{thebibliography}
\end{document}